# WikiTexVC: MediaWiki's native LaTeX to MathML converter for Wikipedia

Johannes Stegmüller[1,2], Moritz Schubotz[1,2],

[1] Zentralblatt Mathematik, FIZ Karlsruhe, Germany ( jstegmuellerthesis@gmail.com)

[2] University of Göttingen, Germany ({last}@gipplab.org)

## ABSTRACT

MediaWiki and Wikipedia authors usually use LaTeX to define mathematical formulas in the wiki text markup. In the Wikimedia ecosystem, these formulas were processed by a long cascade of web services and finally delivered to users' browsers in rendered form for visually readable representation as SVG.

With the latest developments of supporting MathML Core in Chromium-based browsers, MathML continues its path to be a de facto standard markup language for mathematical notation in the web. Conveying formulas in MathML enables semantic annotation and machine readability for extended interpretation of mathematical content, in example for accessibility technologies. With this work, we present, *WikiTexVC* a novel method for validating LaTeX formulas from wiki texts and converting them to MathML, which is directly integrated into MediaWiki. This mitigates the shortcomings of previously used rendering methods in MediaWiki in terms of robustness, maintainability and performance. In addition, there is no need for a multitude of web services running in the background, but processing takes place directly within MediaWiki instances. We validated this method with an extended dataset of over 300k formulas which have been incorporated as automated tests to the MediaWiki continuous integration instances. Furthermore, we conducted an evaluation with 423 formulas, comparing the tree edit distance for produced parse trees to other MathML renderers. Our method has been made available Open Source and can be used on German Wikipedia and is delivered with recent MediaWiki versions. As a practical example of enabling semantic annotations within our method, we present a new macro that adds content to formula disambiguation to facilitate accessibility for visually impaired people.

## KEYWORDS

Math, formulas, MathML , LaTeX , accessibility, blind , MediaWiki , Wikipedia

## 1 INTRODUCTION

The free online encyclopedia Wikipedia has nearly 7 Million articles [1] in the English language. Pages with mathematical formulas are very common on Wikipedia and vast amount of daily users. For example, the Normal Distribution Wiki page was visited by 192,000 unique users [2] in October 2023. The most formulas on Wikipedia are written in LaTeX-format within the markdown describing Wiki pages, the Wiki text. As a result of a search query to the English Wikipedia on August 9, 2023, there are 38,670 pages with the TeX Math tag and 4,097 pages with the alternative HTML Math tag.

MediaWiki is a software to create Wikis and used as the technological foundation for Wikipedia [3] as well as a significant number of other wikis. The MediaWiki stats site[3] counts around 400 entries of such Wikis as of Nov. 2023. To this time, Wikiapiary.com counts 46,380 active sites using MediaWiki. Since MediaWiki is licensed by GNU Public license, it can be used in a wide range of public and some even commercial scenarios for free.

The MediaWiki Math extension is an add-on to the MediaWiki in PHP, providing support for rendering and processing mathematical formulas [4]. The extension supports the *texvc* [11] package of 727 of LaTeX macros used for writing formulas in MediaWikis math environment in wiki text. TexVC is an abbreviation for, *Tex validator and converter*, and throughout MediaWikis' history, several software components have been created to handle the package's macros. This is mostly described in section 3.1.

In the current *SVG rendering* default setting, a TexVC parser as well as the converter components are running inside a dedicated web service called *Mathoid*. This web service communicates with the Math Extension through a messaging server on Wikimedia instances on Wikipedia. In simple MediaWiki deployments, communication occurs directly via REST calls instead.

In both cases, at least running and maintaining one additional web service is required. Mathoid is written in JavaScript using node.js. On Wikipedia, a messaging system called *RESTBase* increases the complexity of the setup and makes maintenance and the implementation of additional features even more difficult. Furthermore, WikiMedia's usage of RESTBase is planned to be discontinued. The adaptation to a new messaging system will require enormous efforts from the current default setup.

Mathematical Markup Language, or MathML is a markup language based on XML for describing mathematical notation. It is a proposed standard which has been drafted by the W3C since 1998 and its goal is to enable the processing and representation of mathematical notations on the web [2]. With the adaptions of MathML Core to Chromium-based browsers [5] in early 2023, support for rendering MathML in the browsers Chrome, Edge, Safari, Firefox, Opera [6] and more has been enabled. Therefore, MathML can be seen as a de facto standard for the web.

In this work, we introduce *WikiTexVC* a solution to integrate LaTeX to MathML parsing natively in MediaWiki and to decouple the translation of the specific TexVC LaTeX package from external translation libraries. Only MathML is delivered to the users by MediaWiki when reading the site, since the browsers are ready to display formulas without SVG. The architecture of WikiTexVC is described in section 3. In our evaluation we investigate WikiTexVC for the criteria of maintenance, robustness and also the performance as well as the quality of conversions. The evaluation and its results can be found in section 4. Furthermore, we introduce a novel accessibility feature within WikiTexVC for the annotation and disambiguation of formulas within screen readers for visually impaired persons.

---

[1] https://de.wikipedia.org/wiki/Wikipedia:Sprachen
[2] https://en.wikipedia.org/wiki/Wikipedia:WikiProject_Mathematics/Popular_pages

[3] https://www.mediawiki.org/wiki/Sites_using_MediaWiki/en
[4] https://www.mediawiki.org/wiki/Extension:Math
[5] https://mathml.igalia.com/news/2023/01/10/mathml-in-chromium-project-is-completed
[6] https://caniuse.com/mathml

The major contributions of this work are the following:

- Introduction of the WikiTexVC Formula validation system and a native translator component from LaTeX/texvc to MathML as an Open Source contribution[7] to the MediaWiki math extension.
- Deployment of these components to MediaWiki and German Wikipedia
- An example implementation of an accessibility feature for visually impaired persons for demonstrating the feasibility of adding new features to WikiTexVC

At the time of writing, native rendering mode can be enabled in the user account settings of German Wikipedia and in MediaWiki instances, further details can be found in the Math Extensions documentation [8].

## 2 RELATED WORKS

### 2.1 MathML

MathML, short for Mathematical Markup Language, is an XML-based language specifically designed for representing and describing mathematical notation, encompassing both its structural layout and content elements. For this, MathML specifies two types of markup elements. One for displaying mathematical expressions, *Presentation markup*. And the other, *Content Markup*, is used to convey mathematical meaning. The idea of content markup is to provide an explicit encoding of the underlying mathematical meaning of an expression, rather than any particular rendering for the expression. [2]

MathML Core, the MathML subset suitable for Browser implementation, which is supported at the time of this writing by major browsers, covers (most of) Presentation Markup, with the focus being the precise details of displaying mathematics in web browsers [4]. For displaying rendered formulas to MediaWiki to users of a Wiki, the usage of MathML core elements, mostly the parts of presentation MathML, is sufficient for translating the LaTeX macros from the *TexVC* package. The MathML specification is iteratively updated and maintained by the W3C Math Working group. The *intent*-attribute is part of the current draft for the upcoming MathML 4.0 standard [9]. It is defined as an accessibility attribute, that helps accessibility technologies (AT) for visually impaired persons to generate accurate audio or braille renderings. This will be further described in our demo application, see section 3.4.

### 2.2 MathJax

MathJax is a JavaScript and, from MathJax version 3, also a typescript library for displaying mathematics typesetting on websites. It is widely distributed on the Internet and on frequently visited websites such as Stack Exchange. Within Mathoid, it is also integrated to Wikipedia for rendering *texvc* LaTeX. It supports various input formats, including LaTeX, MathML, and AsciiMath input formats, as well as HTML with-CSS and scalable vector graphics (SVG) and MathML as output formats. [5].

### 2.3 Mathoid

Mathoid is a JavaScript-based server component, which is used as a web-service in the Wikimedia ecosystem, to convert math input to SVGs. It uses Node.js and phantomJS to run MathJax in a headless browser [12]. For formula validation, Mathoid uses *texvcjs* [1], a JavaScript library which uses parser expression grammar to validate the processed formulas and correct some possible syntax issues, for further processing with MathJax. Wikipedia and larger wikis which are using RESTBase usually are running Mathoid as a web service. RESTBase is a caching and storing API proxy backing the Wikimedia REST API[10]. Wikipedia and MediaWiki up to the current MediaWiki 1.41 use Mathoid based rendering over RESTBase as the default setting for processing formulas.

### 2.4 LaTeXML

LaTeXML[11] is a Perl based software system which converts LaTeX documents into to XML, HTML, MathML and further formats. It was developed for the digital library of mathematical functions[12] (DLMF). One of the design goals, is the lossless conversion from LaTeX to MathML, by preserving both, semantic and presentation cues [9]. In MediaWiki, it can be selected as an option for rendering LaTeX. It can then be used as an external service and requested via HTTP requests. However, it is disabled in Wikimedia deployments such as Wikipedia because there is no way to run Perl-based systems.

### 2.5 The MaRDI Portal

The *MaRDI Portal*[13] is a one-stop access point for mathematical research data which aims to offer comprehensive access to all open research data. It is in active development since late 2021 by Germany's National Research Data Initiative (NFDI). [13]

The portal is based on Wikibase, a series of MediaWiki extensions for realizing knowledge graphs like Wikidata. As of December 2023, the knowledge graph (KG) in the portal contains 12,551 DLMF formulas. In addition, MaRDI strives to present mathematical formulas from various sources in the Wiki texts in MediaWiki and KG of the Portal. WikiTexVC was developed in constant exchange with the MaRDI community, and the MaRDI portal is one of the first applications in which WikiTexVC is used productively.

## 3 SYSTEM OVERVIEW

Figure 1 shows the processing of mathematical formulas in Wiki-TexVC from their occurrence in the wiki text to the rendered form in the HTML-DOM delivered to the browsers of visitors. Formulas are extracted and processed from the Wikipage after page edits

---

from authors. In case an input formula contains additional chemical formula typesetting, it is preprocessed by the *mhchemParser* component, which translates the mhchem typesetting language to regular LaTeX math. This is described in more detail in 3.3. For the depicted example, we choose a fictive formula which contains mathematical and chemical typesetting elements. In the next processing step, the formula is traversing the *texVC* component, which uses a parser expression grammar [15]. In case the input formula is well-formatted, a parse tree is formed which can be used for the translation step to MathML. The parse tree generation is also a formula validation step, if the formula is not formatted according to the grammar, WikiTexVC recognizes the issue in the statement and indicates an error or warnings to the authors. In severe cases, the formula is not further processed. The validation and parse tree creation is described in detail in section 3.1. By creating a structured parse tree, the translation to MathML is enacted by using character mappings and further processing, this is described in section 3.2. After initial MathML translation, a formula can be delivered in the HTML-DOM to the browsers of page visitors. Usually, the formula is cached so that it does not have to be translated multiple times.

## 3.1 Input validation and parse tree creation: TexVC

TexVC is an abbreviation for *Tex validator and converter*. It was initially released as software written in OCAML [15] and later ported as *texvcjs* to JavaScript for the usage of the REST-based rendering pipeline with Mathoid [1]. TexVC validates the correctness of TeX-based inputs for the accurate usage of the currently 727 LaTeX commands [11] supported within the math environment in Wiki text of MediaWiki. In addition, it allows the generation of optimized TeX for further rendering steps with Mathoid.

In TexVC the parser uses a parsing expression grammar (PEG), which is a way of defining a syntax for a language using a set of rules that specify how to match and consume the input text. By parsing a valid TeX-based equation, the PEG produces an abstract syntax tree (AST). This parse tree represents the hierarchical structure and meaning of the TeX-based formulas typically used as input. The AST can then be used for further processing, such as MathML or corrected Tex rendering.

For formula validation, each input expression is checked against a finite set of matching rules defined as regular expressions during the AST creation. When the parser detects text which does not match any of the regexes, the TexVC formula validation process indicates wrong or malicious input within MediaWiki user interface.

A simplified parse tree for the input below can be seen in figure 1.

$$(K+1)^2 = \frac{[\text{Hg}^{2+}][\text{Hg}]}{[\text{Hg}_2{}^{2+}]}$$

The parse tree structure formed by TexVC is similar to the structure of the corresponding MathML for each TeX expression. To use the produced AST as a foundation for native MathML generation, we developed a PHP port of TexVC.

Using this technique only requires the hardware-intensive creation of *one* syntax tree that can be used for both TeX-validation and MathML rendering. In contrast, there is the dedicated creation of the AST in TexVC and within MathJax in the REST-based pipeline.

| TeX-cmd | translation fct. | parameters |
|---------|-----------------|------------|
| ddot | accent | '00A8' |
| tilde | accent | '007E' |
| bar | accent | '00AF' |
| matrix | matrix | |
| array | matrix | |
| pmatrix | matrix | '(', ')' |

**Table 1: TeX to MathML translation mapping table example**

## 3.2 Native MathML generator

In the previous step, TexVC has generated the parse tree with parser expression grammar. The nodes of the tree represent common syntactic expressions in the texvc typesetting language used in the MediaWiki mathematical environments. For example, the *Fun2* node in the parse tree in figure 1 represents a function with two mandatory arguments. The *Curly* node indicates that its child node *Text* is wrapped in curly brackets.

The structure of nodes in the obtained parse tree roughly corresponds to the alignment of MathML elements. For translation of TeX to MathML, we apply a visitor pattern to traverse the AST from its root. In the visitor methods, each TeX command or macro is translated to the corresponding MathML elements and subsequently all necessary transitions and transformations are done.

The translation and transformation steps for each TeX node are usually defined by its arguments. To cover wide language capabilities, we are incorporating mapping tables as they are used in MathJax. These tables map specific node arguments to corresponding translation functions. These functions are usually grouped by the similarity of transformation steps. In table 1 a partial example for such a mapping table can be found. *ddot*, *tilde* and *bar* are parsed in similar ways, so for all of them the "accent" function is used, necessary translation parameters can be directly stored in the translation table for effective software maintenance. In our table, we also include an operator directory which maps TeX-operators, usually recognized as *Literal* nodes by PEG to their corresponding MathML elements. Furthermore, mapping tables for AMS-LaTeX package have been included.

For creating MathML in the translation functions, we implemented 27 presentation MathML version 3 elements from the W3C definition [14].

Since the generated MathML is delivered to the browsers is used for accessibility annotations as well as rendering the visual representation, we choose to generate *presentation markup* in our pipeline. It is used as mathematical notation for a focus on visual display[15]. In contrast to *content markup*, it is supported by the major browser rendering engines, like gecko [16] and WebKit [17].

Through the modular structure with dedicated notations for valid commands in the parse tree, mapping tables and translation functions, security is guaranteed through the validation layer while the software is extensible with further commands and macros.

---

[14]https://www.w3.org/TR/MathML/mathml.html
[15]https://www.w3.org/Math/whatIsMathML.html
[16]https://udn.realityripple.com/docs/Web/MathML/Element
[17]https://trac.webkit.org/wiki/MathML

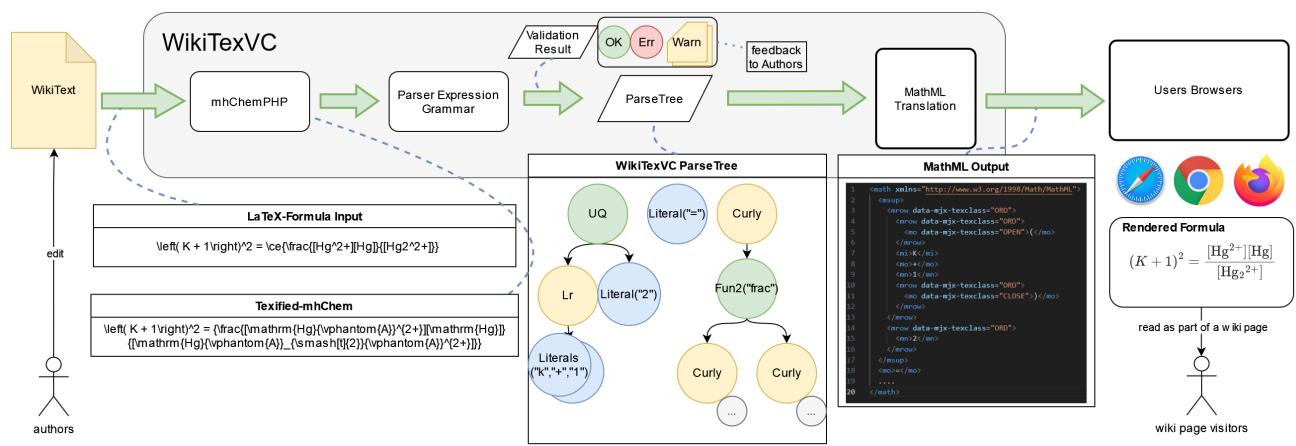

**Figure 1: Overview of components of WikiTexVC, converting Formulas from LaTeX to MathML**

### 3.3 mhchemParser

Chemical formulas are denoted with the *mhchem* typesetting language in MediaWiki [7]. This enables authors to intuitively write chemical formulas (*ce*) as well as physical units (*pu*) in dedicated environments within the math tag.

Consider the following equation written in mhchem:

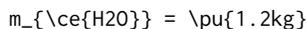

This notation is much simpler and intuitive than writing in LaTeX:

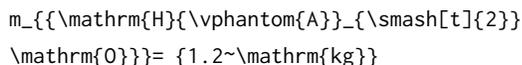

To enable this simplified notation for chemical formulas in our native MathML generation pipeline and for their annotations, we ported mhchemParser [18] from typescript to PHP. To ensure the functionality of the port, which is part of the Math extension, we incorporated 116 tests for validation.

Also, the created software was validated with around 44k tests by Martin Hensel checking the mhchem specification v4.1.2.

In our native MathML generation pipeline, we implemented mhchemParser as a preprocessor to PEG. Each command containing chemical environments is translated to raw LaTeX and afterward the AST is created by PEG.

The mhchemParser requires the introduction of some additional TeX-commands to the grammar. For these and for the conversion of chemical formulas to MathML, we created more than 170 tests within the Math extension.

### 3.4 Accessibility in WikitexVC: Supporting the Intent Attribute

An issue with MathML, as defined in MathML 3, is that it does not convey enough information to provide unambiguous accessibility output in numerous instances. By interpreting formulas denoted in this format, accessibility technologies for visually impaired people, such as screen readers, are not able to produce accurate speech output. In the upcoming version 4 of MathML, the intent attribute is introduced to provide information about the intended meaning of a mathematical expression, for resolving such cases [19]. As an example of the our system's ability to add accessibility features, we present in WikiTexVC a way to annotate LaTeX formulas and create *intent* enhanced MathML output.

*3.4.1 Ambiguity Issue.* In our example for describing the issue, we will use the formula $(0, 5)$. The formula can be interpreted as a point, an open interval, a GCD, a cycle, or an ordered tuple, vector etc. Without further information, the formula in the listing is translated into MathML as seen in listing 3.4.1.

```
<math display='block'>
  <mrow>
    <mo>(</mo>
    <mi>x</mi>
    <mo>,</mo>
    <mi>y</mi>
    <mo>)</mo>
  </mrow>
</math>
```

**Listing 1: formula in MathML**

In the most cases a screen reader, would read MathML from 3.4.1 like this:

*"open paren x comma y, close paren"*

By adding intent-attribute annotations to the MathML in the previous example, screen readers will be able to disambiguate the meaning of the formula. This annotation is usually done directly in the MathML.

```xml
<math display='block'>
  <mrow intent='open-interval($x,$y)'>
    <mo>(</mo>
    <mi arg='x'>x</mi>
    <mo>,</mo>
    <mi arg='y'>y</mi>
    <mo>)</mo>
  </mrow>
</math>
```

**Listing 2: annotated formula in MathML**

By reading the annotated MathML from figure 3.4.1 screen readers can interpret the equation as open interval and read:

*"the open interval from x to y"*

Or, with another annotation:

*"the point x comma y"*

*3.4.2   Solution in WikiTeXVC.* MathML4 [20] defines a basic grammar for the syntax of the intent attribute. Using this grammar as a baseline, we implemented a parser based on PEG, which validates the syntactic correctness of formulas annotated on MediaWiki. In our implemented grammar, which can be seen in the listing 3.4.2, we added some additional attributes like the *structure* attributes and hints for commas in decimals, which are currently in discussion in the W3C intent draft.

```
1  intent := concept-or-literal | number | reference | application |
             structure
2  concept-or-literal := NCName
3  number := '-'? digit+ ( '.' digit+ )?
4  reference := '$' NCName
5  structure := ':' ('common' | 'structure' | 'chemistry' | 'matrix' )
6  application := intent hint? '(' arguments? ')'
7  arguments := intent ( ',' intent )*
8  hint := '@' ( 'prefix' | 'infix' | 'postfix' | 'function' | 'silent'
           | "decimal-comma" | "thousands-comma" )
```

**Listing 3: Implemented grammar for validating intent syntax**

The possible concepts of intent are currently collected and edited collaboratively by W3C in an online spreadsheet [21].

To leverage the full language capabilities of the intent syntax without creating an additional surface language, we implemented functionalities for annotating TeX with unmodified intent and argument syntax.

We introduce a TeX macro which can wrap any Math-TeX expression and mark additional arguments:

```
\intent{<Actual math equation>}
{intent=<intent-syntax> ,arg=<opt additional arguments> }
```

WikiTeXVC implements a *intent* annotation for LaTeX formula. With this annotation macro we can add intent-information:

```
\intent{(x,y)}{intent='open-interval(\$x,\$y)'}
```

Since the the identifiers $x$ and $y$ for the arguments in the formula for open interval and the intent annotation are exactly the same. The MathML output of the intent MathML generator can be seen in listing 3.4.1. If it is necessary to connect differing argument identifiers, a mapping can be specified with the *arg* parameter within the TeX macro.

---

[20] https://www.w3.org/TR/mathml4/
[21] https://shorturl.at/efAF7

---

We incorporated automated tests[22] by extracting around 88 formulas from the W3C intent-examples. Our automated tests checks the validity of intent-grammar in our annotations, furthermore it can export the generated annotated MathML, so that the synthesized speech from the formula can be investigated with accessibility speech tools like MathCat. The baseline testing data and further tools for future processing and evaluation,- as soon as intent is finally defined,- can be found on GitHub [23].

## 4   EVALUATION

For evaluating WikiTexVC, we used the main requirements for a MathML converter on Wikipedia, as stated for Mathoid in 2014 [12]. Since a new rendering method is introduced with WikiTeXVC, we add the evaluation criterion of *structure similarity*. In the beginning of each subsection, the requirement is described concisely. Then we present our evaluation approach and the results. In the following conclusion section, we summarize the results of our evaluation.

### 4.1   Coverage

*Criterion-Desciption:* The converter must support all commands currently used on Wikipedia.

*State:* We implemented an automated test[24] which incorporates all 688 supported LaTeX commands [25] and macros on Wikipedia. Only these commands can be used on Wikipedia, they are verified by the TexVC validation component for the REST-based as well as the native MathML rendering. The automated test is verifying that all commands with their corresponding parameters are rendered correctly as MathML. For a quick structural comparison, we implemented an F-type measure, which checks the created MathML structure against an LaTeXML example. Also, the output can be visually compared by generating an HTML file, which can be inspected in the browser. All measures show that the supported commands are supported by the MathML pipeline and are rendered correctly.

*4.1.1   TexVC Coverage.* To ensure that the functionality of the PHP port of TexVC covers the entire set of the supported LaTeX commands, several automated tests have been incorporated in the MediaWiki-math extension. A comprehensive overview of these tests can be seen in table 2. Due to the test structure, in most cases the number of assertions represents the number of formulas checked.

*4.1.2   Coverage of WikiTeXVC MathML parsing.* The number of possible formula expressions with the 724 TeX commands and macros allows expressing a vast amount of formula. We incorporated several automated tests to make sure the newly generated MathML matches current standards.

---

[22] https://github.com/wikimedia/mediawiki-extensions-Math/blob/4b30c9701a20b5c935c9ffddad071c17dbf71818/tests/phpunit/unit/WikiTeXVC/Intent/intent_mathml_testing_latex_annotated.json
[23] https://github.com/Hyper-Node/AccessibilitySpeechEval/tree/master
[24] https://github.com/wikimedia/mediawiki-extensions-Math/blob/5dd7d8cba34f59abaffce0a3f7585445b7195ccf/tests/phpunit/unit/TexVC/MMLGenerationTexUtilTest.php
[25] https://github.com/wikimedia/mediawiki-extensions-Math/blob/5dd7d8cba34f59abaffce0a3f7585445b7195ccf/tests/phpunit/unit/TexVC/TexUtilRef.json

| Test | Tests | Assertions |
|------|-------|------------|
| All TexVC commands | 40 | 157 |
| Chemical Formulas | 2 | 14287 |
| Real English Wiki formulas | 314 | 936571 |
| Test output correction | 15 +2 | 405 + 8 |

**Table 2: Validating TexVC basic functionality**

As reference for comparing our output, we used Mathoid as it is currently (as of 08.23) deployed on Wikimedia systems. Furthermore, for cases which are not completely covered by Mathoid yet, and to have another output for potential improvements, we used the output of LaTeXML [6] [26]. In figure 2 rendered MathML formulas for a variety of tests for our system in comparison to the reference outputs can be seen.

To have a practical measurement technique in development scenarios, whether the MathML output of our generator matches the reference MathML without performing a hardware-intensive visual comparison, we designed a method which compares the MathML elements within the results and calculates an F-Score based on their similarity. The order of the elements in the results often differs while having very similar visual representations from the rendering. Therefore, we are not considering the order of elements in this measure. Inferred *mrows* and mismatched attributes can be specified to be ignored in the comparison method. For an in-depth structural evaluation for this publication, we used a tree edit distance algorithm, described in subsection 4.6.

- MMLFullCoverageTest: Full coverage test with 424 formulas. It was extracted from a Wikipage for the TexVC validator [27]. We implemented a maintenance script to automatically create MathML references to regenerate the test file for the cases extracted from Wikitext.
- MMLGenerationTexUtilTest: This test with 705 formulas ensures that all covered TexVC commands can be translated to valid MathML, it generates test examples from the definition of supported commands. Commands with arguments have been enriched with realistic argument examples.
- MMLGenerationParserTest: A test with 461 formula examples extracted from Wikipedia.
- MMLRenderTest: A test with in-depth checks for 28 formulas

### 4.2 Scalability

*Criterion-Description:* The load for the converter may vary significantly, since the number of Wikipedia edits heavily depends on the language. Thus, a converter must be applicable for both small and large Wikipedia instances.

*State:* Wikipedia is deployed by the Wikimedia Foundation [28]. This guarantees already high scalability through the utilization of load-balancers, which forward to multiple content delivery networks and server instances. The converter, as direct part of the

Math-Extension, is running directly in numerous instances of application servers on Wikipedia. Also, the native rendering pipeline utilizes caching. Caching information for a formula is stored in the main database (MariaDB) once the wiki page with a currently edited formula is saved. The renderer has to be called only once per formula, in other cases the cache can be used. When there are major updates for the rendering of the parser, the caches can be invalidated. The parser performance is scaling up with the number of application server instances used. Over time of usage, the time fetching formulas is reduced and there is less load on the parser, since more and more formulas are cached.

#### 4.2.1 Robustness.

*Criterion-Desciption:* Bad user input, or numerous concurrent requests, must not lead to permanent failure for the converter.

*State:* All user input is validated by the TexVC-validation grammar before the actual MathML conversion starts. If there is erroneous user input, the grammar recognizes this and prompts a corresponding error output. Since the native MathML renderer is running directly within the MediaWiki application servers, the number of concurrent requests are mitigated by the WMF load balancers. In local and MediaWiki setups, the maximum numbers of requests per instance can be limited by a configuration flag. The native MathML parser itself does not maintain an internal state over the sequential processing of formulas, and therefore can't be compromised by multiple requests.

### 4.3 Speed

*Criterion-Desciption:* Fast conversion is desirable for a good user experience.

*State:* For our speed evaluation, we measure the time from the time a wiki page is accessed with a representative set of formulas until the visitor's web browser retrieves the HTML page information with all the formulas. We use the web-developer tools integrated in Chrome and Firefox to measure the timings. On the English beta cluster Wikimedia instance, we compare the loading times of *MathTestPage* [29] to *MathTestPageNative* [30]. To mitigate latency differences, we calculate an average of three attempts from two different locations.

Wikimedia's deployments of the Math extension utilize caching by default. Formulas are usually rendered when the corresponding wiki page is visited initially by any user. After the initial rendering they are stored in a global cache, the rendering does not have to be processed again for any user. If one or more formula are edited on an existing Wikipage, the edited formulas are rendered and the corresponding cache entries are updated. In order to only take into account the times for generation without pre-generated formulas in the cache, we clear the formula cache before each of our measurements. This is possible through a purge action that was implemented during the development [31] of WikiTexVC.

Since the beta cluster instance of Wikipedia has less computational resources, the processing times are much longer than on the main instance of Wikipedia. The architecture of the deployed ecosystem is the same, therefore the timings can be seen as proportional to the main instance.

| TeX-Input | MathML(LaTeXML) | MathML(Mathoid) | MathML(TexVC) |
|---|---|---|---|
| \sqrt{1-z^3}\! | $\sqrt{1-z^3}$ | $\sqrt{1-z^3}$ | $\sqrt{1-z^3}$ |
| \exp_a b = a^b, \exp b = e^b, 10^m \! | $\exp_a b = a^b, \exp b = e^b, 10^m$ | $\exp_a b = a^b, \exp b = e^b, 10^m$ | $\exp_a b = a^b, \exp b = e^b, 10^m$ |
| \bigoplus, \bigotimes, \bigodot \! | $\bigoplus, \bigotimes, \bigodot$ | $\bigoplus, \bigotimes, \bigodot$ | $\bigoplus, \bigotimes, \bigodot$ |
| \supset, \Supset, \sqsupset \! | $\supset, \Supset, \sqsupset$ | $\supset, \Supset, \sqsupset$ | $\supset, \Supset, \sqsupset$ |
| \sideset{_1^2}{_3^4}\prod_a^b | $\sideset{_1^2}{_3^4}\prod_a^b$ | $\sideset{_1^2}{_3^4}\prod_a^b$ | $\sideset{_1^2}{_3^4}\prod_a^b$ |
| \begin{vmatrix} x & y \\ z & v \end{vmatrix} | $\begin{vmatrix} x & y \\ z & v \end{vmatrix}$ | $\begin{vmatrix} x & y \\ z & v \end{vmatrix}$ | $\begin{vmatrix} x & y \\ z & v \end{vmatrix}$ |
| \left . \frac{A}{B} \right \} \to X | $\frac{A}{B}\} \to X$ | $\frac{A}{B}\} \to X$ | $\frac{A}{B}\} \to X$ |
| x={-b\pm\sqrt{b^2-4ac} \over 2a} | $x = \frac{-b \pm \sqrt{b^2-4ac}}{2a}$ | $x = \frac{-b \pm \sqrt{b^2-4ac}}{2a}$ | $x = \frac{-b \pm \sqrt{b^2-4ac}}{2a}$ |

**Figure 2: Rendered MathML from LaTeXML, Mathoid(MathJax) and TexVC**

|  | **Mathoid** | **WikiTexVC** |
|---|---|---|
| MA-Chrome | 18.25 | 13.55 |
| MA-Firefox | 17.73 | 13.36 |
| LDN-Chrome | 19.01 | 12.41 |
| LDN-Firefox | 18.72 | 12.93 |
| Overall Average | **18.43** | **13.06** |

**Table 3: Measuring the retrieval times of wiki pages with formulas in seconds from Mannheim (MA) and London (LDN)**

The results of our performance evaluation in table 3 show, that in average Mathoid takes 18.43 seconds to load the Wikipage, and WikiTexVC takes 13.06 seconds. This is a difference of 5.37 seconds, and WikiTexVC loads 41.11 % faster. The result can be explained by the fact that WikiTexVC does not cause any latency within the Wikipedia ecosystem.

### 4.4 Maintainability

*Criterion-Desciption:* A new tool for a global site the size of Wikipedia must be able to handle tasks with a large management overhead. Therefore, active development over a long period of time is desirable.

*State:* The native MathML converter is part of the MediaWiki math extension since this is an essential component for delivering Mathematical formulas to Wikipedia, it is maintained by one or several persons. Contributors can upload source code and fixes for the codebase, which is validated through code reviews and a large set of continuous integration checks for running automated tests and static code analysis. New macros can be added in dedicated tables and assigned to specific parsing functions, simplifying maintenance at the programming level.

### 4.5 Accessibility

*Criterion-Description:* Providing accessible content to everyone is one of the key goals of Wikipedia.

*State:* Compared to raw SVG images, MathML delivered to AT like screen readers provides additional information and machine-readable format for synthesizing speech.The major screen readers like VoiceOver, Orca, JAWS, and NVDA support processing MathML. Nevertheless, and as mentioned in the introduction of this work, MathML formulas can be still ambiguous to AT. [8] Therefore, we are introducing a method to add intent annotations within this work, a complete description can be found in section 3.4.

### 4.6 MathML Structure Similarity

*Criterion-Description:* The MathML element structure should match the already established methods on Wikipedia.

*State:* We are using the Robust Tree Edit Distance (RTED) metric to quantify and assess the structural similarities and differences between the MathML representations produced by LaTeXML and WikiTexVC. Since the MathML generated by Mathoid was used on Wikipedia for several years, we use it as a reference and compare the structure of the other systems outputs against it. To detect and measure structural dissimilarities, we use the *Robust Tree Editing Distance* (RTED). RTED is robust to different tree shapes and has a reduced asymptotic complexity, therefore it has a reduced processing time compared to other TED algorithms. [10]

| | |
|---|---|
| Number of Formulas | 423 |
| Overall RTED Mathoid-WikiTexVC | 1217 |
| Overall RTED Mathoid-LaTeXML | 3200 |
| Average RTED Mathoid-WikiTexVC | 2.877 |
| Average RTED Mathoid-LaTeXML | 7.565 |

**Table 4: Comparison of MathML outputs from WikiTexVC to LaTeXML**

In our evaluation, we used the java implementation of RTED v1.1. downloaded from Uni Salzburg [32]. We use a dataset of 423 unique formulas containing practical formula examples and a full coverage of supported macros from *TexVC*. Both, the generated dataset [33] and a runner script for the evaluation can be found online [34].

Table 4 shows the results of our structural similarity evaluation. It shows the accumulated tree edit distance for the evaluation set of 423 formulas. The initial full coverage test in PHP has about four more formulas because we do not consider tests that should test code injection or produce error output. For producing the LaTeXML output and Mathoid references, we use the deployed implementations used on Wikipedia as of December 2023.

In its default setting, the LaTeXML output, unlike Mathoid and WikiTexVC, contains the *annotation* element, which allows computer algebra systems or accessibility systems to interpret the meaning of a formula. To obtain accurate results, this element is removed from the LaTeXML in this evaluation. Since both the MathML of LaTeXML and Mathoid are surrounded by a *semantics* element, we add this element to the WikiTexVC for this evaluation.

The results show that the structure of WikiTexVC outputs is much more similar to Mathoid than LaTeXML. LaTeXML needs 1983 additional changes to WikiTexVC to reflect the Mathoid MathML structure. On average, 4,688 additional edits are required per formula. This result is expected because WikiTexVC was designed to produce structurally similar output to Mathoid, and in MathML there are several ways to represent a formula structurally. We can hereby observe that WikiTexVC can achieve the necessary structural similarity to the Mathoid established on Wikipedia

## 5 CONCLUSION

With our publication and deployment of WikiTexVC we offer a way to display mathematics natively on Wikipedia and other wikis. The maintenance effort for adjustments in the package-specific language texvc is significantly lower than in Mathoid. Our solution eliminates the maintenance overhead of routing requests through RESTBase and maintaining two codebases. Wiki solutions generally do not require the provision of additional web services because formula rendering is done by default. Additionally, our processing speed evaluation shows that our system requires less time to render formulas and therefore less CPU time and energy. Through our structure evaluation, we can guarantee a similar shape to the currently used Mathoid rendering.

However, during development and community feedback, we discovered inconsistencies in browser modification, particularly in rendering for Chromium. Once intend is specified and introduced as standard, we would like to produce a more sophisticated intent language implementation which introduces multiple macros for simplified intent user input.

## ACKNOWLEDGMENTS

We thank the Wikimedia Foundation for ongoing support in deployment of the WikiTexVC component and fixing numerous issues in the continuous integration. Furthermore, we thank Martin Hensel for support and code review regarding the mhchem parser port. We thank the W3C Math WG for their continuous feedback and support for understanding the intent-attribute. Our research was supported by the German Research Foundation (DFG grant no.: 460135501).

**Listing 4: Use the following BibTeX code to cite this article**

```bibtex
@inproceedings{Stegmueller2024,
  author = {{Stegm\"{u}ller Johannes, Schubotz Moritz}},
  title = {WikiTexVC: MediaWiki's native LaTeX to MathML converter for
Wikipedia},
  booktitle = {arXiv Preprint, January 2024}
}
```
**references.bib**